\documentclass[10pt,sigplan,letterpaper,twocolumn]{article}
\usepackage[10pt,nocopyright]{sigmin}

\usepackage{csquotes}
\usepackage{listings}
\usepackage{color}
\usepackage{epsfig,makecell}                                      
\usepackage[hyphens]{url}                                         
\usepackage[breaklinks,colorlinks]{hyperref}                      
\usepackage[usenames,dvipsnames]{xcolor}                          
\usepackage{colortbl}
\hypersetup{citecolor=blue,linkcolor=blue}                        
\usepackage{amsmath,amsopn,amssymb,amsthm}                        
\usepackage{thmtools}
\usepackage[toc,page]{appendix}
\usepackage{subcaption}                                           
\usepackage{endnotes,microtype,xspace,fancyvrb,multirow} 
\usepackage{booktabs}                                             
\usepackage{array,underscore,relsize}                             
\usepackage[T1]{fontenc}                                          
\usepackage{times}                                                
\usepackage{fancyhdr,lastpage}                                    
\usepackage{enumitem}                                             
\usepackage[labelfont=bf,font=normal,skip=1pt]{caption}           
\usepackage[export]{adjustbox}                                    
\usepackage{breakurl}                                             
\usepackage{pifont}                                               
\usepackage{tablefootnote}                                        
\usepackage[many]{tcolorbox}
\usepackage[linesnumbered,ruled,noend,scleft]{algorithm2e}                     

\SetNlSty{large}{}{:}

\SetCommentSty{mycommfont}
\SetArgSty{textnormal}
\SetKwComment{Comment}{\textcolor{blue}{$\triangleleft$\ }}{}
\SetKwProg{Fn}{Function}{:}{}
\newcommand{\pluseq}{\mathrel{+}=}
\newcommand{\minusseq}{\mathrel{-}=}

\usepackage{amssymb}
\usepackage[normalem]{ulem}
\usepackage{changepage}   

\usepackage[hang,flushmargin]{footmisc}
\usepackage{textcomp}
\usepackage{graphicx}

\usepackage[compact]{titlesec}
\titleformat*{\section}{\large\bfseries}
\titleformat*{\subsection}{\normalsize\bfseries}
\titleformat*{\subsubsection}{\normalsize\bfseries}

\hypersetup{
  colorlinks,
  linkcolor={red!50!black},
  citecolor={blue!50!black},
  urlcolor={blue!80!black}
}

\newcommand{\sys}{{\textsc{KRCore}}}

\newcommand{\lite}{\textsc{LITE}}

\def\ie{i.e.,~}
\def\eg{e.g.,~}

\newcommand{\blue}[1]{\textcolor{blue}{#1}}

\newcommand{\TODO}[1]{\textcolor{brown}{TODO: #1}}

\usepackage{tikz}

\usepackage{balance}
\usepackage{authblk}

\begin{document}

\title{\Large \bf {\sys}: A Microsecond-scale RDMA Control Plane for Elastic Computing}

\author[1,2]{Xingda Wei}
\author[1]{Fangming Lu}
\author[1,2]{Rong Chen\thanks{Rong Chen is the corresponding author (\url{rongchen@sjtu.edu.cn})}}
\author[1] {Haibo Chen}
\vspace{-100pt}
\affil[1]{Institute of Parallel and Distributed Systems, SEIEE, Shanghai Jiao Tong University}
\affil[2]{Shanghai AI Laboratory}

\date{}
\maketitle

\frenchspacing

\begin{abstract}
We present {\sys}, an RDMA library with a microsecond-scale control plane 
on commodity RDMA hardware for elastic computing. 
{\sys} can establish a full-fledged RDMA connection within 10${\mu}$s
(hundreds or thousands of times faster than verbs),
while only maintaining a (small) fixed-sized connection metadata at each node, 
regardless of the cluster scale.
The key ideas include virtualizing pre-initialized kernel-space RDMA connections 
instead of creating one from scratch,
and retrofitting advanced RDMA dynamic connected transport with static transport 
for both low connection overhead and high networking speed. 
Under load spikes, 
{\sys} can shorten the worker bootstrap time of an existing disaggregated key-value store 
(namely RACE Hashing) by 83\%.
In serverless computing (namely Fn), 
{\sys} can also reduce the latency for transferring data through RDMA by 99\%. 
\end{abstract}

\section{Introduction}
\label{sec:intro}

The desire for high resource utilization has led to the development of elastic applications
such as disaggregated storage systems~\cite{DBLP:conf/usenix/TsaiSZ20, facebook/dis,racehashing}. 
Elasticity provides a quick increase or decrease of computing resources (\eg{processors or containers})
based on application demands.
Since the resources are dynamically launched and destroyed, 
minimizing the control path overheads---including process startup and creating network connections---is vital to applications,
especially those with ephemeral execution time.
Elastic applications typically have networking requirements. 
For instance, computing nodes in a disaggregated storage system
access the data stored at the storage nodes across the network. 

\begin{figure}[!t]
    \centering\includegraphics[scale=0.9]{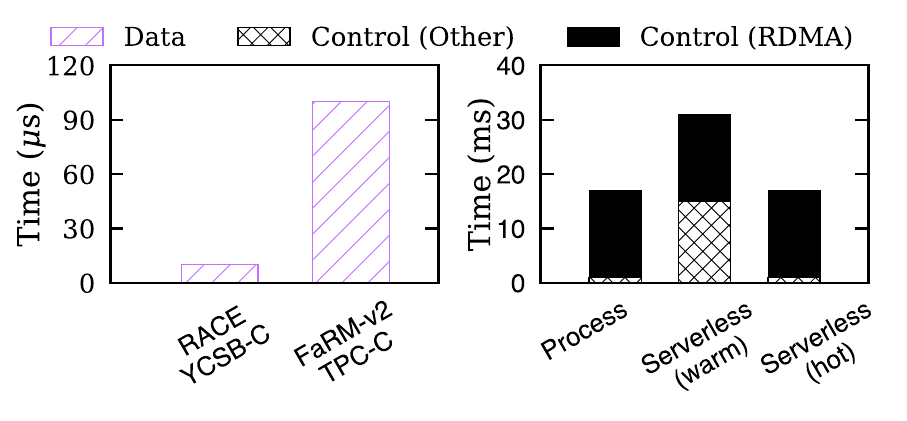} \\[1pt]
    \begin{minipage}{1\linewidth}
    \caption{\small{
        (a) The execution time (\textbf{Data}) of typical elastic RDMA-enabled applications, 
		and (b) the breakdown of control path costs. 
        RACE~\cite{racehashing} is a disaggregated key-value store.  
        FaRM-v2~\cite{farmv2} is a database that can accelerate serverless transactions~\cite{DBLP:conf/osdi/ZhangCCAL20}. 
        YCSB-C~\cite{ycsb} and TPC-C~\cite{tpcc} are representative benchmarks for each system.
        The serverless platform evaluated is Fn~\cite{fn}. 
    }}
    \label{fig:motiv-app}
    \end{minipage} \\[-15pt]
    \end{figure}

RDMA is a fast networking feature widely adopted in datacenters~\cite{tsai2017lite,DBLP:conf/sigcomm/GuoWDSYPL16, farm}. 
Unfortunately, RDMA has a slow control path:
the latency of creating an RDMA connection (15.7ms)
is 15,700X higher than its data path operation (see Figure~\ref{fig:motiv-app}(b)). 
As the latency of typical RDMA-enabled applications that require elasticity 
has reached to microsecond-scale (see Figure~\ref{fig:motiv-app}(a)), 
this high connection time may significantly decrease the application efficiency, 
\eg{increasing latency when expanding resources to handle load spikes}. 
The cost is challenging to reduce 
because it not only includes software data structure initialization costs 
but also involves extensive hardware resource configurations, 
as RDMA offloads network processing to the network card (\textsection{\ref{sec:cost-user}}).

A common approach to avoiding the control path cost 
is to cache connections and 
share them with different applications. 
However, user-space RDMA connections can not be directly shared 
by different applications,
because each app has its own exclusive driver data structure
 and dedicated hardware resources. 
Nevertheless, sharing a kernel-space RDMA connection is possible 
since applications share the same kernel ({\lite}~\cite{tsai2017lite}). 
However, {\lite} has performance and resource inefficiency issues (\textsection{\ref{sec:cost-kernel}}) 
in elastic computing,
because it doesn't target this scenario.
First, it still pays the initialization cost under cache misses.
Second, caching all RDMA connections to all nodes is resource inefficient (\eg{taking several GBs of memory}), 
especially when a production RDMA-capable cluster has reached a scale of more than 10,000 nodes~\cite{xrdma}. 
Finally, sharing RDMA connections complicates the preservation of the low-level verbs interfaces, 
which is important to apply RDMA-aware optimizations~\cite{racehashing,xstore,
dragojevic2015nocompromises,drtmh,herd,DBLP:conf/usenix/KaliaKA16}. 
{\lite} only provides a high-level API. 

We continue the line of reusing connections to boost the RDMA control path,
and further overcome the issues mentioned above.
We present \textbf{{\sys}}, a networking library 
with an ultra-fast control plane. 
{\sys} can establish a full-fledged RDMA-capable connection 
within 10 $\mu$s, 
only 0.05\% and 0.22\% of the verbs and {\lite} 
under cache misses, respectively. 
More importantly, 
{\sys} only needs a small amount of fixed-sized memory for the connection pool (\eg{64MB}),
irrelevant to the cluster scale.
Finally, {\sys} supports low-level RDMA interfaces
compatible with existing RDMA-aware optimizations.

Supporting such a fast control plane
seems to contradict our promise of a small fixed-sized connection pool.  
To achieve this, {\sys} makes a key innovation:
we \emph{retrofit} a less-studied 
yet widely supported advanced RDMA hardware feature---\emph{dynamic connected transport} (DCT)~\cite{dct}---\emph{to the kernel}. 
DCT allows a single RDMA connection to communicate with different hosts. 
Its connection and re-connection are offloaded to the hardware 
and thus, are extremely fast (less than 1$\mu$s). 
Our observation is that 
when virtualizing an established kernel-space DCT connection to different applications, 
they no longer pay the control path cost and memory consumption of ordinary RDMA connections.

In designing {\sys}, we found virtualizing DCT with a low-level API brings several new challenges, 
and we propose several techniques to address them (\textsection{\ref{sec:challenge}}).
First, DCT requires querying a piece of metadata to establish a new connection. 
Using RPC can not achieve a stable and low latency.
Further, RPC needs extra CPU resources to handle DCT-related queries. 
Observing the small memory footprint of DCT metadata, 
we propose an architecture that deploys RDMA-based key-value stores 
to offload the metadata queries to one-sided RDMA READ (\textsection{\ref{sec:data-structure}}). 
Second, DCT has a lower data path performance than normal RDMA transport (RC) due to its dynamic connecting feature. 
The performance is mostly affected when a node keeps a long-term communication with another. 
Therefore,
we introduce a hybrid connection pool that retains a few RC connections connected to 
frequently communicated nodes to improve the overall performance. 
{\sys} further adopts a transfer protocol that can  
transparently switch a virtualized connection from DCT to RC (\textsection{\ref{sec:qp-migrate}}). 
Finally, 
we propose algorithms to safely virtualize a shared physical QP to multiple applications 
with a low-level API (\textsection{\ref{sec:api-data}}). 

We implement {\sys} as a loadable Linux kernel module in Rust. 
We also extended an existing kernel-space RDMA driver (mlnx-ofed-4.9) to 
bring DCT to the kernel.
To the best of our knowledge, {\sys} is the first 
to achieve a microsecond-scale RDMA control plane. 
Although {\sys} is a general-purpose RDMA library,
it really shines with elastic computing applications. 
Our experiments demonstrated that 
{\sys} can reduce the computing node startup time of 
a state-of-the-art production RDMA-enabled disaggregated key-value store (RACE~\cite{racehashing}) by 83\%, 
from 1.4s to 244ms (\textsection{\ref{sec:eval-race}}). 
For serverless computing---another popular elastic application, 
{\sys} can shorten the data transfer time over RDMA by 99\%, from 33.3ms to 0.12$\mu$s
(\textsection{\ref{sec:eval-serverless}}). 

Our source code and experiments are available at 
{\small{\url{https://github.com/SJTU-IPADS/krcore-artifacts}}}. 

\section{Background and Motivation}
\label{sec:bg}

\subsection{The case for fast control path in elastic computing}
\label{sec:context}

{\sys} targets systems that require elasticity: 
the ability to automatically scale according to application demands. 
One such case is disaggregated storage systems 
where the computing nodes and storage nodes are separated and connected by the network~\cite{DBLP:conf/usenix/TsaiSZ20,facebook/dis,racehashing}. 
Under high loads, the system can dynamically add computing nodes for better performance:
and they need to establish connections to the storage nodes on-the-fly.
Another important case is serverless computing~\cite{DBLP:journals/corr/abs-1902-03383}
where the platforms instantaneously launch short-lived tasks with containers\footnote{\footnotesize{Serverless 
platforms may use virtual machine (VM)s to run tasks, which is not the focus of our paper. 
\textsection{\ref{sec:dislim}} discusses how {\sys} can apply to VMs.}}.
The launch time typically includes network connections~\cite{DBLP:conf/cloud/ThomasAVP20}.

Unlike long-running tasks (\eg{web servers}), 
the control path (\eg{network creation}) is typically on the critical path of elastic applications. 
For example, 
before executing the application code,
a serverless function that issues database transactions 
must first establish network connections to remote storage nodes~\cite{DBLP:conf/osdi/ZhangCCAL20, DBLP:conf/sosp/JiaW21}.
With RDMA, 
the transaction latency has reached 10-100${\mu}s$~\cite{dragojevic2015nocompromises,drtmh}. 
Reducing the control path costs---including launching a container and creating network connections---is 
therefore vital to the end-to-end execution time or tail latency of elastic applications (see Figure~\ref{fig:motiv-app}). 

Much research has focused on reducing other control path costs, 
\eg{the container launch time to about 10ms~\cite{sockatc} and even sub-millisecond~\cite{du2020catalyzer}.
However, 
only a few considered accelerating network connection creation~\cite{DBLP:conf/cloud/ThomasAVP20}, 
especially for RDMA. 
The control path of RDMA is indeed 
several orders of magnitude slower than its data path (\eg{22ms vs. 2$\mu$s in \textsection{\ref{sec:bg-issues}}}).
It is also orders of magnitude slower than the execution time of common elastic RDMA-enabled applications,
or other control path costs (see Figure~\ref{fig:motiv-app}).

\subsection{RDMA and queue pair (QP)}
\label{sec:bg-rdma}

RDMA is a high bandwidth and low latency networking feature
widely adopted in modern datacenters~\cite{tsai2017lite,DBLP:conf/sigcomm/GuoWDSYPL16}.
It has two well-known primitives:
\emph{two-sided} provides a message passing primitive while
\emph{one-sided} provides a remote memory abstraction---the
RDMA-capable network card (RNIC) can directly read/write server memory 
in a CPU-bypassing way.

Although RDMA is commonly used in the user-space, 
the kernel adopts the same \emph{verbs} API (verbs), 
which exposes network connections as queue pairs (QPs).
Each QP has a {send queue} (\emph{sq}), a {completion queue} (\emph{comp\_queue}),
and a {receive queue} (\emph{recv\_queue}). 
Both primitives follow a similar execution flow. 
To send a request (or a batch of requests), 
the CPU uses {\texttt{post\_send}} to post it (or them) to the \emph{send queue}. 
If the request is marked as \emph{signaled}, the completion can be polled from the \emph{completion queue} via \texttt{poll\_cq}. 
For two-sided primitive, the CPU can further receive messages with \texttt{poll\_cq} over the \emph{receive queue}. 
Before receiving, 
one should use \texttt{post\_recv} to post message buffers to the QP.
Note that the CPU needs to register memory through \texttt{reg\_mr} to give RNIC memory access permissions. 

QP has several kinds of transport each with different capabilities.
We focus on improving the control path performance of 
\emph{reliable connected} QP (RCQP),
as it is the most commonly used one that 
supports both RDMA primitives and is reliable. 

\begin{figure}[!t]
    \includegraphics[scale=1.1,left]{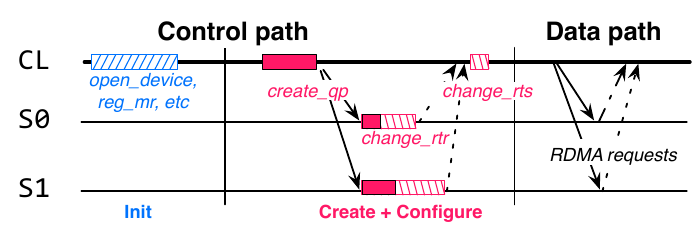} \\[2pt]
    \begin{minipage}{1\linewidth}
    \caption{\small{The execution flow of a client (\textbf{CL}) communicating
    with two nodes (\textbf{S0} and \textbf{S1}) using user-space verbs.   
    \texttt{change\_rtr} changes the QP to ready to receive status while 
    \texttt{change\_rts} changes the QP to ready to send status. 
    }}
    \label{fig:bg-rdma-user} 
    \end{minipage}  \\[-10pt]
    \end{figure}

\begin{figure}[!ht]
\vspace{-2mm}
\centering
\includegraphics[scale=0.9]{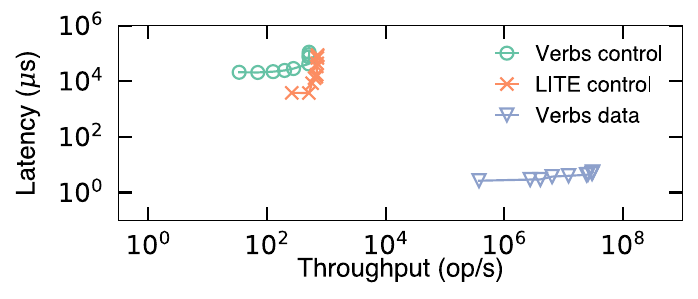} \\[-3pt]
\centering
\includegraphics[scale=0.9]{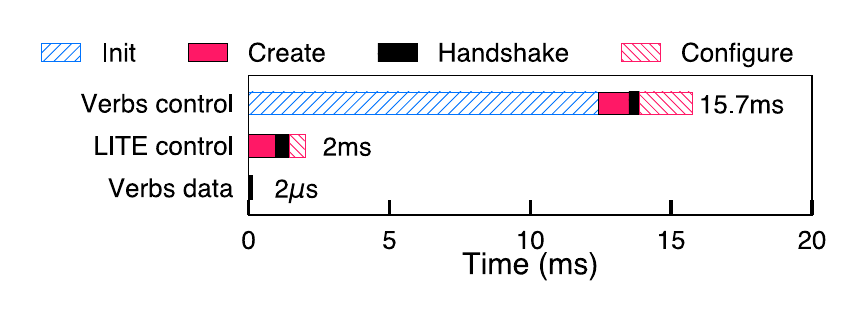} \\[-2pt]
\begin{minipage}{1\linewidth}
\caption{\small{(a) Huge performance gap btw. RDMA's control path and 
data path (issuing 8B READ) when connecting and communicating with one node. 
(b) A breakdown of RDMA control path time.}}
\label{fig:bg-connect-thpt-latency}
\end{minipage} \\[-15pt]
\end{figure}  

\subsection{Analysis of RDMA control path costs}
\label{sec:bg-issues}

\subsubsection{User-space control path costs}
\label{sec:cost-user}

Consider the example in Figure~\ref{fig:bg-rdma-user} 
where a client sends RDMA requests to two nodes.
The control path includes first initializing the driver context (\textbf{Init})\footnote{\footnotesize{Including 
creating the protection domain and registering the memory.}},
creating the QPs (\textbf{Create}), 
exchanging the QP information to the remote peer with a \textbf{handshake} protocol  
and configuring the QPs to ready states (\textbf{Configure}). 
Figure~\ref{fig:bg-connect-thpt-latency}(a) reports its latency, 
which is 7,850X higher than the data path (Verbs control vs. Verbs data). 

\paragraph{Issue: High hardware setup cost.}
To quantify the costs in detail, 
Figure~\ref{fig:bg-connect-thpt-latency} breakdowns the control path time. 
We carefully optimize the connection handshake with RDMA's connectionless datagram~\cite{fasst},
which is orders of magnitude faster than using TCP/UDP.
Contradicting the common wisdom,
exchanging the connection information through the network (\textbf{Handshake}) 
is \textbf{not} the dominant factor:
\textbf{Handshake} only contributes 2.4\% of the total time. 
The cost is dominated by communicating with the RNIC hardware for the connection setups. 
Consider the \texttt{create\_qp} in \textbf{Create}: 
we found 87\% of the \texttt{create\_qp} time (361$\mu$s vs. 413$\mu$s) is waiting for the RNIC 
to create the hardware queues. 

\subsubsection{Existing kernel-space solution is insufficient}
\label{sec:cost-kernel}

{\lite}~\cite{tsai2017lite} is the only kernel-space RDMA solution
and is the closest to our work. 
It provides high-level remote memory \emph{read}, \emph{write} and \emph{RPC} interfaces 
over the low-level verbs API (\textsection{\ref{sec:bg-rdma}}).
{\lite} maintains an in-kernel connection pool 
that caches RCQPs connected to all nodes,
which 
avoids the user-space \textbf{Init} (Figure~\ref{fig:bg-rdma-user}) costs 
because applications share the same kernel-space driver data structures. 
However, it still has the following \textbf{issues}
for elastic applications:

\vspace{2pt}
\paragraph{Issue\#1: High cost connecting to a new node.}
If the RCQP of the target node is not cached, 
{\lite} must follow the same \textbf{Create} and \textbf{Configure} as user-space RDMA, 
\eg{S1 in Figure~\ref{fig:bg-rdma-kernel},
which are non-trivial (2ms for each connection).
Note that we have carefully optimized {\lite}'s control path: 
{\lite} originally adopts a centralized cluster manager to create connections, 
which can only establish tens of QPs per second.  
We optimize it with a decentralized connection scheme 
using RDMA's connectionless datagram. 
The optimization achieves a 2ms per-connection latency 
and 712 QPs/second per node throughput (Figure~\ref{fig:bg-connect-thpt-latency}), 
bottlenecked by the RNIC (see \textsection{\ref{sec:cost-user}}). 

\begin{figure}[!t]
    \vspace{-5pt}
    \includegraphics[scale=1.1,left]{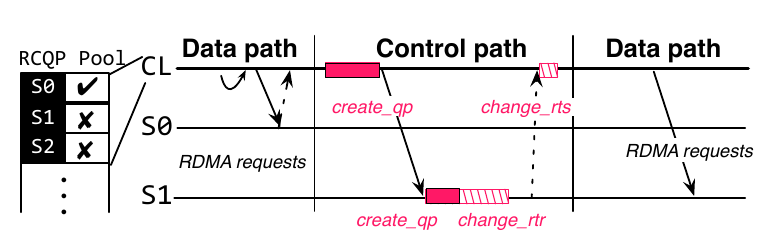} \\[2pt]
    \begin{minipage}{1\linewidth}
    \caption{\small{The execution flow of a client (\textbf{CL}) communicating with 
    two nodes (\textbf{S0} and \textbf{S1}) with the kernel-space RDMA assuming that 
    \textbf{CL} has cached a QP to \textbf{S0} in its connection pool. 
    }}
    \label{fig:bg-rdma-kernel} 
    \end{minipage} \\[-10pt]
    \end{figure}

\interfootnotelinepenalty=10000

\vspace{2pt}
\paragraph{Issue\#2: Huge memory consumption.}
Caching RCQPs connected to all other nodes can mitigate Issue\#1. 
However, this strategy has huge per-machine memory consumption 
since the number of RCQPs needed scales linearly with the cluster size.
In {\lite}, each QP consumes at least 159KB memory\footnote{\footnotesize{{It configures 
the QP with 292 \emph{sq} and 257 \emph{comp\_queue} entries, a common setup in RDMA-based systems.
Each \emph{sq} entry takes 448B while \emph{cq} takes 64B. 
The driver would further round queues to fit the hardware granularity.}}}, 
excluding the message buffers and receive queues (may share between different QPs via shared receive queue).
Therefore, 
{\lite} would consume at least 1.52 GB memory per node for fast connection 
on a modern RDMA-capable cluster with more than 10,000 nodes\cite{DBLP:conf/nsdi/GaoLTXZPLWLYFZL21}}.

\paragraph{Issue\#3: Inflexible interface. }
{\lite} exposes a high-level RDMA API (\eg{a synchronous remote memory read}), 
which simplifies sharing the same QP to different applications. 
However, it is inflexible to apply RDMA-aware optimizations widely adopted in the literature~\cite{racehashing,xstore,dragojevic2015nocompromises,drtmh,herd,DBLP:conf/usenix/KaliaKA16}, 
\eg{sending different read/write requests within a batch asynchronously}. 
To utilize these optimizations, 
applications need verbs low-level API (\textsection{\ref{sec:bg-rdma}}).
Unfortunately, directly executing the low-level API on a shared QP 
can easily corrupt the QP states (see \textsection{\ref{sec:challenge}}),
and interrupt application running.
We carefully design the QP virtualization algorithms to 
correctly virtualize a shared QP with verbs's low-level API (\textsection{\ref{sec:api-data}}). 

\section{Approach and Overview}
\label{sec:overview}

\paragraph{Opportunity: \emph{advanced RDMA transport (DCT). }}    
\label{sec:dct}
Dynamically Connected Transport (DCT)~\cite{dct} is an advanced RDMA feature 
widely supported in commodity RNICs (\eg{from Mellanox Connect-IB~\cite{connectib} to ConnectX-7~\cite{connectx7}). 
DCT preserves the functionalities of RC and
further supports dynamic connecting:  
a DCT QP (DCQP) can communicate to different nodes without user-initiated connections:
RNIC can create DCT connections on-the-fly
by piggybacking control plane messages 
with data plane ones.
Since the connections are only processed in the hardware, 
DCT re-connection is extremely fast: 
our measured overhead is less than 1$\mu$s.
When using DCQPs, 
the host only needs to specify the 
target node's RDMA address and its DCT metadata 
(\ie{DCT number and DCT key}) in each request.

\paragraph{Basic approach: \emph{virtualized kernel-space DCQP.}}
The goal is to achieve an ultra-fast control plane for the applications.
Our basic approach is to 
virtualize kernel-space DCQPs (as VQPs) to user-space applications. 
The observation is that DCT naturally addresses 
the {costly creation overhead (\textbf{Issue\#1})}
and the {huge memory consumption (\textbf{Issue\#2})} 
of RCQPs (\textsection{\ref{sec:cost-kernel}}).
A kernel-space solution further mitigates the user-space 
driver loading costs (\textsection{\ref{sec:cost-user}}). 

VQP also supports low-level RDMA interfaces (\eg{\texttt{ibv\_post\_send}})
with the necessary extended API suitable for elastic computing  (\textsection{\ref{sec:api}}).
Therefore, users can flexibly apply existing RDMA-aware optimizations~\cite{herd,DBLP:conf/usenix/KaliaKA16, drtmh} 
 (\textbf{Issue \#3} in \textsection{\ref{sec:cost-kernel}}). 
Note that different VQPs can share the same physical QP in the kernel.
Nevertheless, {\sys} provides an exclusively owned QP abstraction 
to the applications. 

\subsection{Challenges and solutions}
\label{sec:challenge}


\paragraph{C\#1. Efficient DCT metadata query.}
DCQP needs to query the DCT metadata before sending requests. 
Specifically, to allow communicating with DCT, 
the server must first create a \emph{DCT target}
identified by a key and number (DCT metadata).
Afterward,
the clients can piggyback the metadata in their requests 
to communicate with the created target.

A viable solution is to 
send an RPC to the target node to query the metadata
using RDMA's connectionless datagram (UD)\footnote{\footnotesize{It only supports two-sided RDMA. }},
 which prevents control plane costs as UD is connectionless.  
However, it is inefficient in performance and CPU usage. 
First, the latency of RPC may vibrate to tens of milliseconds 
due to the scheduling and queuing overhead of the CPU. 
Second, {\sys} must deploy extra kernel threads 
to handle the queries.

\begin{adjustwidth}{0pt}{}
    \underline{\emph{Solution: RDMA-based meta server. }}
    \; We replicate the DCT metadata at a few global {meta servers} 
    backed by RDMA-enabled key-value stores (KVS)~\cite{racehashing,DBLP:conf/sosp/WeiSCCC15,xstore,farm},
    meaning each node can query it with one-sided RDMA bypassing the CPU.
    To support one-sided RDMA while preventing QP over-provisions, 
    {\sys} only maintains a few RCQPs connected to nearby meta servers.
    Replicating the DCT metadata is practical because it is small: 
    12B is sufficient for one node to handle all requests from others. 
\end{adjustwidth}

\begin{figure}[!t]
    \vspace{-3pt}
    \centering
    \includegraphics[scale=1.0,left]{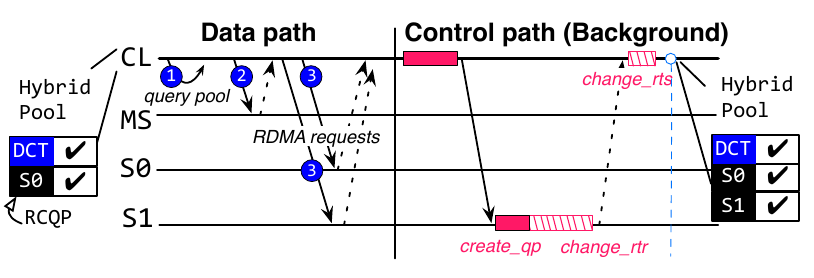} \\[1pt]
    \begin{minipage}{1\linewidth}
    \caption{\small{
        The execution flow of a client (\textbf{CL}) communicating with two nodes (\textbf{S0} and \textbf{S1}) with {\sys}. 
        \textbf{MS}: meta server. 
        Note that {\sys} always put the hardware control path (\ie{creating RCQPs}) in the background.
        }}
    \label{fig:overview}
    \end{minipage} \\[0pt]
    \end{figure}

\paragraph{C\#2. Performance issues of DCT.}
DCT is slower than RC in peak throughput and may incur 
high tail latency due to re-connection (\textsection{\ref{sec:eval-data}}).
The performance is mostly affected 
when a node frequently sends requests to the same node. 

\begin{adjustwidth}{0pt}{}
\underline{\emph{Solution: virtualized hybrid QP.}}
\; {\sys} manages a hybrid QP pool that stores both RC and DC QPs.
A VQP can transparently switch between DC and RCQP (\textsection{\ref{sec:qp-migrate}}), 
allowing us to create RCQPs in the background
on-the-fly without exposing the 
creations overhead to the applications.
\end{adjustwidth}

\paragraph{C\#3. QP state protection.}
If we directly forward the VQP request (from \texttt{ibv\_post\_send}) 
from different applications
to the (same) shared physical QP,
QP's physical states can easily be corrupted 
due to malformed requests or queue overflow,
because verbs API assumes an exclusively owned QP.
Bringing the QP back to a normal state is costly because 
it requires reconfiguration (the \textbf{Configure} in Figure~\ref{fig:bg-connect-thpt-latency} (b)). 

\begin{adjustwidth}{0pt}{}
\underline{\emph{Solution: pre-check.}}
\; {\sys} carefully checks the physical queue capacity and request integrity 
before forwarding the requests to the physical QP. 
The overhead of these checks is negligible as they only involve simple calculations.
Thus, we can avoid QP corruption while 
preserving the RDMA-aware optimizations (\textsection{\ref{sec:api-data}})
of using low-level interfaces.
\end{adjustwidth}

\subsection{Execution flow and architecture} 
\label{sec:arch}

\paragraph{Execution flow.}
Applications can use {\sys} to create RDMA-capable connections in a few microseconds. 
Figure~\ref{fig:overview} presents its execution flow when communicating to two nodes. 
First, we find available RCQPs in the hybrid pool (\blue{\ding{182}}). 
If exists (S0), we directly virtualize it. 
Otherwise (S1),  we choose a DCQP and 
fetch the target node's DCT metadata (\blue{\ding{183}}) accordingly. 
Finally, we virtualize the selected QP so that 
the client can send RDMA requests with them (\blue{\ding{184}}). 

To increase the likelihood of hitting RCQPs, 
{\sys} analyzes the host's networking patterns 
and creates RCQPs in the background (\eg{to S1}). 

\begin{figure}[!t]
	\vspace{-2mm}
    \centering\includegraphics[scale=1.28]{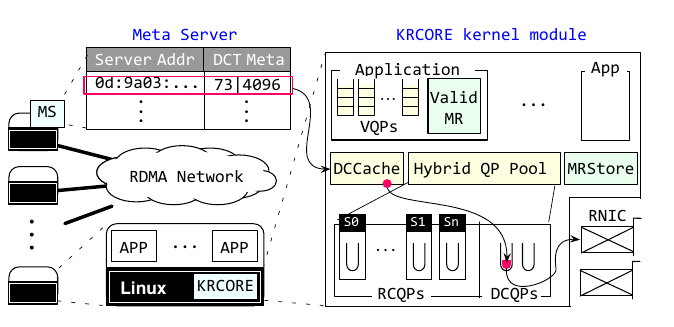} \\[-0pt]
    \begin{minipage}{1\linewidth}
    \caption{\small{An overview of {\sys} architecture.}}
    \label{fig:arch}
    \end{minipage} \\[5pt]
    \end{figure}

\paragraph{Architecture.}
Figure~\ref{fig:arch} presents the {\sys} library architecture.  
On each node, 
{\sys} is a loadable Linux kernel module
hosting per-application (\eg{VQP}) and per-node (\eg{Hybrid QP Pool}) data structures (\textsection{\ref{sec:data-structure}}). 
{\sys} also deploys meta servers (MS) 
on a few nodes to facilitate DCT metadata lookup.
These servers are backed by DrTM-KV~\cite{DBLP:conf/sosp/WeiSCCC15}---a state-of-the-art RDMA-enabled KVS---to accelerate the metadata lookup. 
The metadata is broadcasted by each machine during its boot time.

\section{Detailed Design}
\label{sec:design}

\subsection{Programming interface of {\sys}}
\label{sec:api}

To simplify application development and porting, 
it is important to keep backward compatibility between {\sys} and verbs, 
the de facto standard for using RDMA. 
In principle, {\sys} can provide the same interface with verbs
similar to existing work (\ie{Freeflow~\cite{freeflow}}). 
However, verbs is not designed for elastic computing 
and may bring inflexibility or under-utilization of {\sys}. 
Therefore, we propose an extended API based on verbs
inspired by Demikernel~\cite{DBLP:conf/sosp/ZhangRPONLMLSJP21}, 
as shown in Figure~\ref{fig:api}. 
Specifically, {\sys} introduces a new type of QP (\emph{VQP}) 
with the following new primitives:

\begin{figure}[!t]
    \centering
    \includegraphics[scale=0.37]{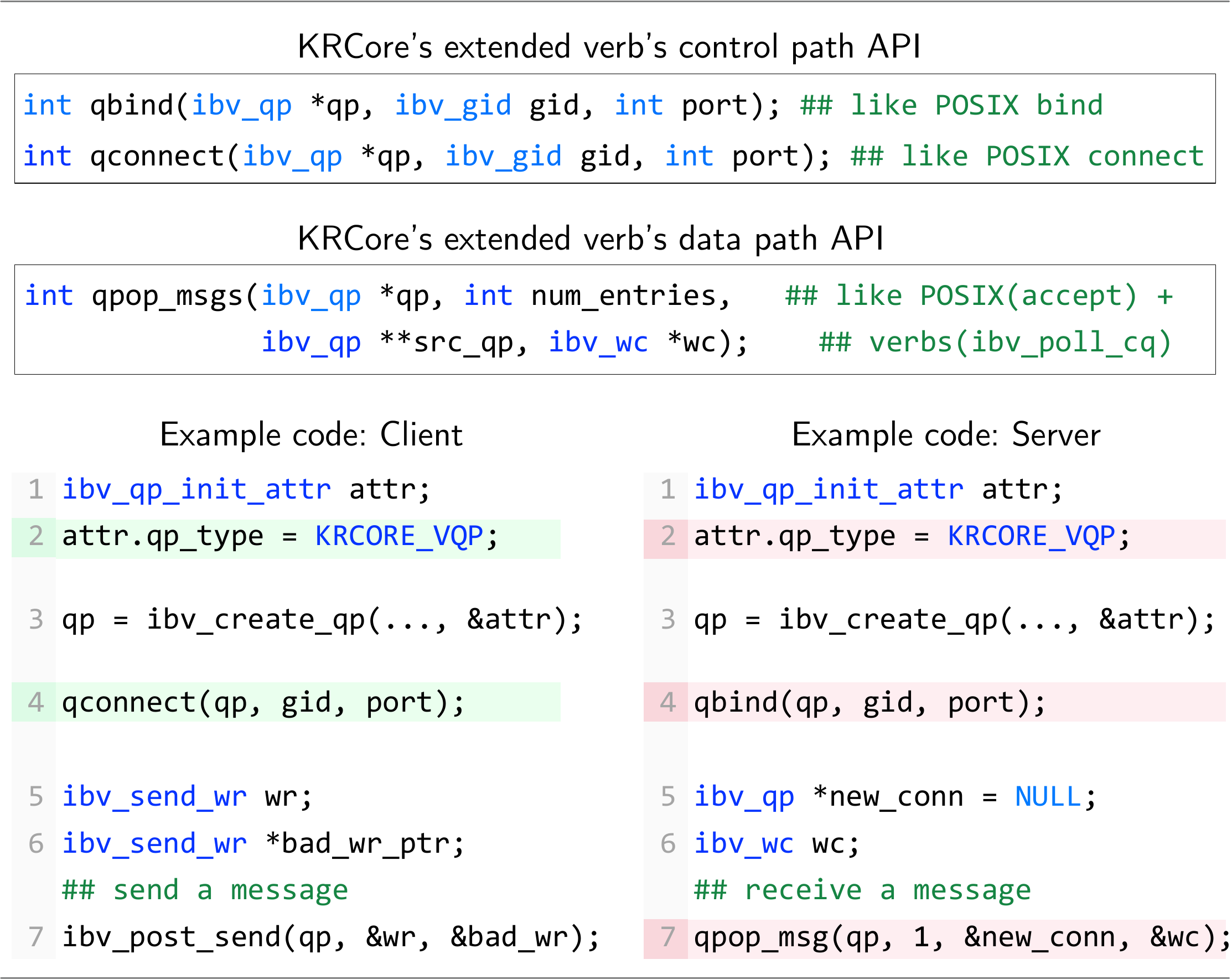} \\[10pt]
    \begin{minipage}{1\linewidth}
    \caption{\small{
        The {\sys} extended API atop of verbs
        and a simplified use case.        
        Lines in {\color[HTML]{DDFBE6}{$\blacksquare$}} and {\color[HTML]{F9D7DC}{$\blacksquare$}}  
        are extended code for the client and server, respectively. 
        Applications can also use the verb's data path call (\eg{\texttt{ibv\_post\_send}})
        to issue RDMA requests with {\sys}. 
    }}
    \label{fig:api}
    \end{minipage} \\[-10pt]
    \end{figure}

\paragraph{\texttt{qconnect} and \texttt{qbind}.}
The verbs API has no method for `connect' commonly found in networking libraries. 
Therefore, developers have to implement and optimize RDMA connection setups themselves. 
We provide a \texttt{qconnect} API to abstract the fast connection provided by {\sys}.
Specifically, 
after calling \texttt{qconnect} on a VQP to a remote host (identified by the RDMA address (gid) and a port),
the VQP can issue one-sided and two-sided requests to it. 
Note that remote end must bind to the address using \texttt{qbind} beforehand
so that the sender can issue two-sided requests,
similar to POSIX \texttt{bind}.

\paragraph{\texttt{qpop\_msgs}.}  
RCQPs are one-to-one connected---meaning the server must know how many 
clients may connect. 
This is unhandy for elastic applications because
clients can dynamically connect to a server. 
Therefore, {\sys} VQP is many-to-one:
after binding to an address, 
a VQP can dynamically accept new connections when receiving messages: 
\texttt{qpop\_msgs} will return a list of $(src\_qp, message)$ pairs, 
where the $src\_qp$ is a VQP connected to the corresponding sender of the $message$.

\vspace{1mm}
Besides the extended API,
{\sys} also supports common verbs data path API, \eg{
\texttt{ibv\_post\_send}, \texttt{ibv\_post\_recv} and \texttt{ibv\_poll\_cq} 
(see \textsection{\ref{sec:bg-rdma}})}. 
Figure~\ref{fig:api} showcases a simplified code example 
of sending a message from a client to a server with VQP.
At the client, it can use \texttt{KRCORE_VQP} as a marker to create a VQP. 
After successfully connecting the VQP with \texttt{qconnect}, 
the client can call \texttt{ibv\_post\_send} to send the message.

Note that the VQP has the semantic as RCQP---meaning that 
they have reliability guarantees and support all RDMA operations 
(with various low-level optimizations). 

\subsection{Data structures}
\label{sec:data-structure}

\paragraph{Hybrid QP pool.}
Each VQP (\textsection{\ref{sec:api}}) is backed by a kernel-space virtual QP
that has an identifier, a reference to a physical QP
and virtualized counterparts of RDMA queues (see \textsection{\ref{sec:bg-rdma}}). 
The physical QP is selected from a hybrid QP pool with both DCQPs and RCQPs.
The DCQPs are statically initialized upon boot time and 
RCQPs are created on-the-fly.

In principle, 
the pool only needs one DCQP to
handle all the RDMA requests of the host.
However, 
only using one DCQP introduces extra latency when 
sending concurrent requests to different servers.
Specifically, 
if two requests targeting different hosts go over the same DCQP, 
the second must wait for an additional reconnection 
before RNIC can process it. 
This can be mitigated by increasing the DCQP pool size
since reconnections can run concurrently. 
Yet,
the best choice of the pool size depends on the hardware setting (\textsection{\ref{sec:eval-data}}).
On our platform, we choose 8 DCQPs in the pool.

To further prevent lock contention~\cite{fasst},
we divide the pool on a per-CPU basis:
Each VQP only virtualizes QPs from its local CPU's pool.
This strategy is optimized for cases when each QP 
is exclusively used by one thread, 
a common pattern in RDMA applications~\cite{wukong, xstore,fasst, DBLP:conf/nsdi/GaoLTXZPLWLYFZL21,DBLP:conf/usenix/LuSCL17}. 
In case of thread migrations, 
{\sys} also re-virtualizes QPs in the background 
with a transparent QP transfer protocol (\textsection{\ref{sec:qp-migrate}}).

\paragraph{Meta Server.}
\label{sec:metaserver}
For steady and low-latency DCT metadata query, 
we replicate all the nodes' metadata
at a few global meta servers backed by DrTM-KV~\cite{DBLP:conf/sosp/WeiSCCC15}, 
a state-of-the-art RDMA-enabled KVS. 
Note that replicating all the DCT meta at one server is practical 
because they are extremely small (\eg{17KB for a 1,000-server cluster}). 

The meta server stores a mapping between the RDMA address (key) 
and its corresponding DCT number and key (value). 
These key-value pairs can be queried via DrTM-KV with a few one-sided RDMA READs.
Since sending one-sided requests also requires RDMA connections, 
each node pre-connects to nearby meta servers (\eg{one in the same rack}) with RCQPs during boot time and thus, 
it can find the DCT metadata of a given server in several microseconds even under high load.

\paragraph{Optimization: DCCache.}
Observing that the DCT metadata is extremely small (12B), 
each node further caches them locally 
to save network round-trips querying the meta server.
The metadata is suitable for caching because 
they are only invalidated when the corresponding host is down.

\paragraph{ValidMR and MRStore.}
To safely virtualize a physical QP to multiple VQPs, 
{\sys} additionally checks the validity of remote memory accesses 
to prevent QP state corruption (\textsection{\ref{sec:api-data}}).
These checks were originally done by the RNIC 
using the information stored in the NIC cache. 
Thus, we should also record them in {\sys}.
We additionally bookkeep the registered memory regions (MR)s in ValidMR,
which is also implemented with DrTM-KV. 
After the bookkeeping, 
{\sys} can query the local/remote ValidMRs 
to check the local/remote memory regions' validity.

Like DCCache, 
we also cache the checked remote MR locally (in MRStore) to avoid extra round-trips. 
However, caching remote MRs may introduce consistency problems:
unlike long-lived DCT metadata, 
MRs are managed by the applications and can be de-registered on-the-fly.
To this end,
{\sys} adopts a lease-based lightweight invalidation scheme:
the cached MRs are periodically (\eg{1 second}) flushed. 
Upon de-registration, {\sys} waits for this period before freeing the MR. 

\subsection{Control path operations}
\label{sec:api-ctrl}

\setlength{\textfloatsep}{5pt}
\begin{algorithm}[!t]
    \DontPrintSemicolon
    \footnotesize
    \caption{\small{VQP creation and connection}}  \label{alg:connection}
    \label{alg:create-connect}

    \SetAlgoLined
\SetKwFunction{FMain}{vqp_create}
\Fn{\FMain{$Q$}}{ 
    $Q.id \gets$ allocate a free identifier \label{alg:create:id}\;
    $Q.comp\_queue \gets$ allocate a software queue \label{alg:create:queue}\;
    $Q.recv\_queue \gets$ allocate a software queue\;
    $Q.qp \gets$ NULL  ~~\Comment{Updated by qconnect} \label{alg:connect:create} 
}

\vspace{1mm} 
\SetKwFunction{FConnect}{vqp_connect}
\Fn{\FConnect{$Q$, $addr$}} {
    \If{$Q.qp ==$ NULL} {
        \uIf{$addr$ in $HybridQPPool.RC$} { \label{alg:connect:rc}
            $Q.qp \gets $ select in $HybridQPPool.RC[addr]$ \label{alg:connect:rc-end}}
        \Else{  
            $Q.qp \gets$ select in $HybridQPPool.DC$ \label{alg:connect:dc}\;

            \If{$addr$ not in $DCCache$} { \label{alg:connect:dc:meta}
                $meta \gets$ query nearby connected MetaServer  \label{alg:connect:dc:query} \;
                add $meta$ to $DCCache$ \;
                $Q.dct\_meta \gets meta$ \;
            } \label{alg:connect:dc:meta:end}
        }   
    }    
}
\end{algorithm}

{\sys} reuses initialized QPs upon VQP connection and creation, 
whose simplified pseudocode 
executed in the {\sys} kernel
is shown in Algorithm\ref{alg:create-connect}.

\texttt{vqp_create} initializes the basic data structures of VQP---mainly
allocating the software send and completion queues in the kernel. 
The physical QP assignment is delayed to the VQP connection (line~\ref{alg:connect:create})
because we are unaware of the remote target during creation. 

\texttt{vqp_connect} connects a VQP to a remote end by 
assigning a pre-initialized kernel-space QP (either RCQP or DCQP) to it.
Given the remote $addr$,
it first checks whether an RCQP is available in the HybridQPPool (line~\ref{alg:connect:rc}). 
If so, we choose an available QP and assign it to $Q.qp$ (line~\ref{alg:connect:rc-end}). 
Otherwise, we select a DCQP (line~\ref{alg:connect:dc}).    
Note that all DCQPs in the pool are available 
because {\sys} can virtualize one physical QP to multiple VQPs (\textsection{\ref{sec:api-data}).

When assigning a DCQP to VQP, 
we need to fetch the remote end's DCT metadata
(line~\ref{alg:connect:dc:meta}$-$\ref{alg:connect:dc:meta:end}) 
if the metadata is not cached in the $DCCache$. 
We issue one-sided RDMA READs to the $MetaServer$ 
to query it (line~\ref{alg:connect:dc:query}).

\paragraph{Background RCQP creations.} 
To increase the likelihood of hitting an RCQP in the pool, 
{\sys} maintains background routines to sample frequently communicated nodes,
create RCQPs for frequently communicated ones in the $HybridQP Pool$
and reclaim rarely used RCQPs.
Currently, we choose a simple LRU strategy for the reclamation.

\paragraph{Other control path operations.}
Besides VQP creation and connection,
other control path operations (\eg{memory registration, MR}) 
have a straightforward implementation: 
we forward them to the corresponding verbs API
and record the results in {\sys}. 
If necessary, we will also return the virtual handler of 
the recorded results to the user. 
Due to space limitations, we omit a detailed description. 

\subsection{Data path operations}
\label{sec:api-data}

As we have mentioned in \textsection{\ref{sec:challenge}}, 
a key challenge in virtualizing a physical QP to multiple VQPs 
is preventing shared QP state corruption.
Specifically, we must consider:

\vspace{-8pt}
\begin{enumerate}[leftmargin=*]
    \itemsep0em    
    \setlength{\itemindent}{0em}
    \item \textbf{Detecting malformed request.}  
    An incorrect operation code or an invalid memory reference would transit 
    a QP into error states.  Since an error states QP cannot handle any RDMA requests,
    we must filter out malformed requests before posting them to the physical QP. 

    \item \textbf{Preventing NIC queue overflow. }
    The physical QP has a limited queue capacity. 
    If the user overflows a QP, the QP will also enter an error state. 
    Preventing queue overflow is challenging under sharing 
    because it can overflow even if all the shared users correctly 
    avoid the queue overflows.
    
    \vspace{-2pt}
    The queue can be cleared via explicit signaling and polling.
    Nevertheless, 
    we should poll as little as possible 
    because they have overheads~\cite{herd}.     

    \item \textbf{Dispatching completion events.}
    The polled results of a physical QP can be from different VQPs. 
    Therefore, 
    we must correctly dispatch them to the targets, 
    \ie{software queues of VQPs}.     
\end{enumerate} 
\vspace{-4pt}

\setlength{\textfloatsep}{5pt}
\begin{algorithm}[!t]
\DontPrintSemicolon
\footnotesize
\caption{\small{kernel handler of post\_send and poll\_cq}}  \label{alg:data}
\label{alg:qpush}
    
\SetAlgoLined
        
\SetKwFunction{FPush}{post\_send\_virtualized}
\Fn{\FPush{$Q$, $wr\_list$}} { \Comment{$wr\_list$: the RDMA requests list}

        \Comment{Assumption: the size of wr\_list is smaller than $Q.qp.sq.max\_depth$ and $Q.qp.cq.max\_depth$}        

        \While(){$Q.qp.sq.max\_depth$ - $Q.qp.uncomp\_cnt$ $<$ $wr\_list.length$}{ \label{alg:push:overflow:start}
            \texttt{poll\_inner($Q$)}\;\label{alg:push:overflow:end}
        }
        
        $unsignaled\_cnt \gets 0$\label{alg:push:overflow:local}\;

        \For{$req$ in $wr\_list$} { 

            \If{$req$ has invalid MR or invalid Op} { \label{alg:push:check} 
                \Return{$Error$}
            }

            \uIf(){$req$ is $signaled$} { 
                $Q.comp\_queue.add({Not Ready, req.wr\_id})$\label{alg:push:dispatch}\;
                $req.wr\_id \gets $ encode the pointer of $Q$ and $(unsignaled\_cnt + 1)$\label{alg:push:overflow:encode}\;
                $unsignaled\_cnt \gets 0$}
            \Else {
                $unsignaled\_cnt$ $\pluseq$ $1$\label{alg:push:overflow:add}\;
            }
            $Q.qp.uncomp\_cnt$ $\pluseq$ $1$\;
        }

        \If{$last\_req$ in $wr\_list$ is not $signaled$}  { \label{alg:push:mark:start}
            mark $last\_req$ as $signaled$\;
            $last\_req.wr\_id \gets $ encode NULL and $(unsignaled\_cnt + 1)$\label{alg:push:mark:end}\;
        }

        \Return{\texttt{post\_send($Q.qp$, $wr\_list$)}\label{alg:push:post}}\; 

}

\vspace{1mm}      
\SetKwFunction{FPopInner}{poll\_inner}
\Fn{\FPopInner{$Q$}} {
    $wc \gets $ \texttt{poll\_cq($Q.qp.cq$)}\label{alg:push:overflow:poll}\;
    \If{ $wc$ is ready } {
        $VQ$, $comp\_cnt \gets$ decode $wc.wr\_id$ \label{alg:pop:inner:start}\;
        $Q.qp.uncomp\_cnt \minusseq comp\_cnt$\label{alg:push:overflow:decrease}\;
        \If{$VQ$ is not NULL} { 
            $VQ.comp\_queue.head()[0] = Ready$\label{alg:pop:inner:end}\;
        }
    }    
}

\vspace{1mm}
\SetKwFunction{FPop}{poll\_cq\_virtualized}
\Fn{\FPop{$Q$}} {
    \texttt{poll\_inner($Q$)}\;
    \If {$Q.comp\_queue.has\_head()$ \textbf{and} $Q.comp\_queue.head()[0]$ is ready} { 
        $user\_wr\_id \gets Q.comp\_queue.pop()[1]$\;
        \Return{READY, $user\_wr\_id$}\label{alg:pop:end}
    }
    \Return{NULL, 0}
}
\end{algorithm}  

\noindent
To this end, 
{\sys} will (1) check the request integrity before posting it to a shared QP; 
(2) inject necessary polls to the physical QP and 
(3) encode the VQP information in the request's $wr\_id$---that will
be returned upon request completion---to help the dispatch. 
Specifically, 
{\sys} executes \texttt{post\_send\_virtualized} and \texttt{poll\_cq\_virtualized} 
after the user calls \texttt{ibv\_post\_send} and \texttt{ibv\_poll_cq}, respectively. 
Algorithm~\ref{alg:qpush} shows their simplified pseudocode.
For simplicity,
we assume the request list ($wr\_list$) depth is smaller than the QP capacity, 
which can be achieved by segmenting the request list before posting it.  

\paragraph{\texttt{post\_send\_virtualized}.}
It first clears the physical QP's send and completion queues to prevent overflows (line~\ref{alg:push:overflow:start}$-$\ref{alg:push:overflow:end})
via polling the physical completion queue (line~\ref{alg:push:overflow:poll}). 
Polling is tricky when considering unsignaled requests---the 
requests that don't generate completion events. 
Their entries are freed until a later signaled request is polled. 
Thus, 
we must track how many requests a signaled one is responsible to clear (line~\ref{alg:push:overflow:local} and line~\ref{alg:push:overflow:add}),
and encode the number in $wr\_id$ (line~\ref{alg:push:overflow:encode}). 
Therefore, after polling a completion we can determine the left spaces of queues (line~\ref{alg:push:overflow:decrease}). 
Further, if the last request is unsignaled, we signal it 
(line~\ref{alg:push:mark:start}$-$\ref{alg:push:mark:end}).

For each request, we also check 
whether it is malformed (line \ref{alg:push:check})
and record the dispatch information for the signaled ones 
(line~\ref{alg:push:dispatch}$-$\ref{alg:push:overflow:encode}). 
Finally, we can safely post these requests to the physical QP (line~\ref{alg:push:post}).

For two-sided primitive, {\sys} must additionally notify the receiver the sender information.
Otherwise, the receiver cannot create proper connections 
in \texttt{qpop\_msgs}. 
Hence, we piggyback the sender's address in the message header 
(omitted in the algorithm). 

\paragraph{\texttt{poll\_cq\_virtualized}.} 
It first calls ~\texttt{poll\_inner} to poll the physical QP events
and dispatch the events to the proper VQPs
according to the information recorded in the $wr\_id$ 
(lines~\ref{alg:pop:inner:start}$-$\ref{alg:pop:inner:end}). 
After the dispatch, it can check whether the virtualized QP
has a completion event.
{\sys} examines the head of the virtualized comp\_queue
and returns the head's $wr\_id$ to the application if the head exists.

Due to space reasons, we briefly describe other operations: 

\paragraph{\texttt{ibv\_post\_recv}.}
This function registers the buffers to the VQP
by recording them in the virtualized recv\_queue.

\paragraph{\texttt{qpop\_msgs}.}
It polls the physical QP's $recv\_queue$ and dispatches the received messages,
similar to \texttt{poll\_inner}. 
To hold in-coming messages, 
we pre-post message buffers to physical QP
before virtualizing it to the applications.
The challenge of pre-post is that the {\sys} doesn't know the exact payloads 
of the incoming messages. 
For now, we assume the pre-posted buffers can always hold the incoming message.
\textsection{\ref{sec:two-sided}} 
will describe how we cope with out-of-bound messages in detail. 
After receiving a message, we will check its destination VQP 
and copy it the user-registered buffer (from \texttt{ibv\_post\_recv}).

Besides receiving messages, 
\texttt{qpop\_msgs} also creates a VQP connected to the sender
(\textsection{\ref{sec:api}}). 
The creation and connection follow the control path operations
discussed in \textsection{\ref{sec:api-ctrl}}. 
To prevent the DCT metadata query, 
we further piggyback the metadata in the message header.
Thus, \texttt{qpop\_msgs} doesn't involve additional networking requests.

\subsection{Zero-copy protocol for two-sided operations} 
\label{sec:two-sided}

The basic \texttt{qpop\_msgs}  (\textsection{\ref{sec:api-data}}) has two issues. 
First, it incurs extra memory copies.
Though the copy overhead is negligible for small messages (\eg{less than 1KB}), 
it is non-trivial for the large ones (\eg{see results in Figure~\ref{fig:dc-meta} (b)}). 
Second, it cannot receive messages with payloads larger than the pre-posted buffers.

To this end, we adopt a zero-copy protocol to overcome the above issues.
Intuitively, 
for large or out-of-bound messages, 
the receiver will use one-sided RDMA READ to 
read them to the user-registered buffers,
inspired by existing RDMA-enabled RPC frameworks~\cite{rfp,DBLP:conf/nsdi/GaoLTXZPLWLYFZL21}.
Specifically, 
if the payload is larger than the kernel's registered buffer, 
the sender will first send a small message 
containing the destination VQP ID, 
the source message address and its size.
The receiver can then use one-sided RDMA READ to read the message directly
to the user-registered buffer in a zero-copy way. 
The cost of sending an additional message is trivial for large messages
because the network transfer will dominate the time.

\subsection{Physical QP transfer protocol}
\label{sec:qp-migrate}

{\sys} supports seamlessly changing the physical QP virtualized by a VQP to another.
The challenge of doing so is 
how to preserve the RCQP's FIFO property~\cite{ib} of the VQP during transfer,
\ie{after a request completes, all its previous requests are finished}. 

To ensure FIFO, upon the transfer starts, 
we first post a fake signaled RDMA request to the source QP and wait for its completion before the change. 
Meanwhile, 
we also notify the remote peers to transfer their physical QP.
Otherwise, the VQP can no longer receive the remote end's message. 
For correctness, we must wait for the remote acknowledgments 
before changing the physical QP at the sender. 

\section{Evaluation}
\label{sec:eval}

We aim to answer the following questions during evaluations:
\begin{enumerate}[leftmargin=*, topsep=2pt,itemsep=2pt,partopsep=0pt, parsep=0pt]
\item How fast is the {\sys} control plane (\textsection\ref{sec:eval-control})? 
\item What are the costs to the data plane (\textsection\ref{sec:eval-data})?
\item How RDMA-aware applications that require elasticity can benefit from {\sys} (\textsection\ref{sec:eval-app})? 
\end{enumerate}

\paragraph{Implementation.}
We implement {\sys} from scratch as a loadable Linux kernel (4.15) module,
which has more than {10,000 LoC} Rust code.
It exports system calls via {$ioctl$} without modifying the kernel.
To simplify user-kernel interactions,
we further implement a 100 LoC C shim library atop ioctl
to provide the interfaces described in \textsection{\ref{sec:api}}.
Finally, we port DCT to the kernel-space RDMA driver 
by adding around 250 LoC C code to the {mlnx-ofed-4.9} driver: 
DCT is currently only implemented in the user-space RDMA drivers.

\paragraph{Testbed setup. }
We conduct experiments on a local rack-scale RDMA-capable cluster 
with ten nodes. 
Each node has two 12-core Intel Xeon E5-2650 v4 processors, 128GB DRAM
and one ConnectX-4 MCX455A 100Gbps InfiniBand RNIC. 
All nodes are connected to a Mellanox SB7890 100Gbps InfiniBand Switch.
Without explicit mention,
we deploy one meta server for {\sys}. 

\paragraph{Comparing targets. }
We compare {\sys} with user-space verbs (\textbf{verbs}) and {\lite}\footnote{\footnotesize{\burl{https://github.com/WukLab/LITE}}}.
Original {\lite} has an unoptimized control plane: 
it uses a centralized cluster manager to establish connections between servers 
and can only connect tens of RCQPs per second.
Therefore, we further optimize it 
by enabling a decentralized QP connection scheme
via RDMA's unreliable datagram (UD).
Our optimized version can achieve an optimal kernel-space RDMA 
control plane performance---it is now only bottlenecked by the hardware limits.
\textsection{\ref{sec:eval-control}} will describe this in more detail. 
Note that our optimization leaves the {\lite} data plane unchanged.

\subsection{Control path performance}
\label{sec:eval-control}

The evaluations for the control path focus on creating and connecting 
RDMA connections.
The costs of the other operations in {\sys} (and verbs) are 
typically much smaller. 
For example, registering 4MB memory 
only takes 1.4$\mu$s in {\sys}. 
Therefore, we omit their results. 

We use two synthetic workloads (single and full-mesh connection establishment) 
to evaluate the control path performance.
The connection pool and DCCache of {\sys} are cleared before the evaluations.
Otherwise, {\sys} only has system call overheads and is extremely small (0.9$\mu$s).

\begin{figure}[!t]
        \begin{minipage}{.48\linewidth}    
                \hspace*{-2mm}
                \includegraphics[scale=1]{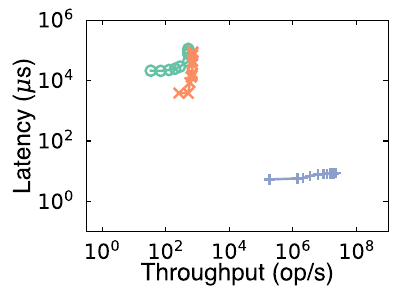} 
                \end{minipage}
                \begin{minipage}{.48\linewidth}    
                \hspace*{-2.5mm}
                \includegraphics[scale=1]{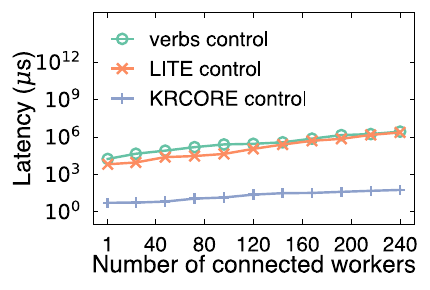} 
                \end{minipage}  \\[2pt]
                \begin{minipage}{1\linewidth}
                    \caption{\small{  
                        The \texttt{qconnect} performance of {\sys} when using DCQP with DCT metadata uncached. 
                        (a) Connecting to a single server, and (b) establishing connections in a full-mesh fashion. 
                        }}                      
        \label{fig:connect-perf}
        \end{minipage} \\[5pt]
        \end{figure}

\vspace{-1.5mm}
\paragraph{Single-connection establishment performance.}
We first evaluated the latency and throughput of establishing 
a single RDMA-capable connection to one server
w.r.t. the number of clients.
Figure~\ref{fig:connect-perf} (a) reports the throughput-latency graph 
when increasing the number of clients from 1 to 240. 
From the figure we can see that 
{\sys} can have several orders of magnitude better performance 
than verbs and {\lite}. 
At one client, 
{\sys} can establish a connection in 5.4$\mu$s, 
while verbs and {\lite} take 15.7ms and 2ms, respectively. 
The performance gain of {\sys} comes from replacing 
the costly RDMA control path operations (analyzed in \textsection{\ref{sec:cost-user}} and \textsection{\ref{sec:cost-kernel}} in detail)
with fast RDMA data path operations,
\ie{two one-sided RDMA READs to the meta server}.
For {\lite}, it saves the driver loading cost 
but still needs to create and configure QP on its control path. 
At 240 clients,
{\sys} can handle 22 million (M) connections per second,
while verbs and {\lite} can only establish 712 RCQPs per second.
They are both bottlenecked by the server creating hardware resources, 
while {\sys} always reuses existing ones to prevent these overheads.


\paragraph{Full-mesh connection establishment performance. }    
Besides establishing a single connection, 
creating full-mesh connections at a set of workers is common in elastic applications,
 \eg{burst-parallel serverless workloads~\cite{DBLP:conf/cloud/ThomasAVP20}}. 
Specifically, each worker should connect to the others and vice versa. 
Figure~\ref{fig:connect-perf} (b) presents the full-mesh performance
by varying the number of involved workers.
In general, {\sys} can reduce 99\% of the full-mesh creation time regardless 
of the worker number, 
thanks to the orders of magnitude faster single-connection establishment performance (see Figure~\ref{fig:connect-perf} (a)).
For example, {\sys} connected 240 workers in 81 $\mu$s, 
while verbs and {\lite} used 2.7 secs and 2.3 secs, respectively. 
These results suggest that {\sys} can handle complex control path 
operations well. 

\begin{figure}[!t]
    \begin{minipage}{.48\linewidth}    
            \hspace*{-2mm}
            \vspace{1mm}
            \includegraphics[scale=1]{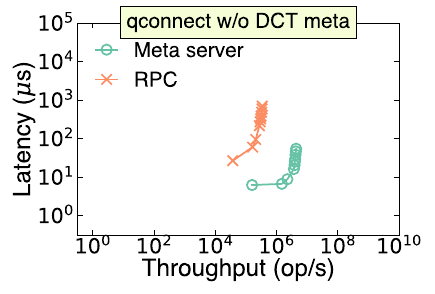} 
            \end{minipage}
            \begin{minipage}{.48\linewidth}    
            \hspace*{-1mm}
            \includegraphics[scale=1]{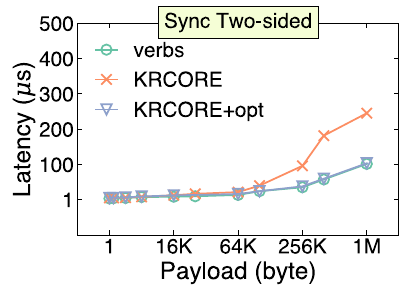} 
            \end{minipage}  \\[2pt]
            \begin{minipage}{1\linewidth}
                \caption{\small{  
                (a) Performance comparisons of different DCT meta query methods, and (b)
                the effects of zero-copy protocol ({\sys}+opt) of {\sys} two-sided operations. 
        }}                      
    \label{fig:dc-meta}
    \end{minipage} \\[0pt]
    \end{figure}

\paragraph{Benefit of the meta server. }
A key design choice of {\sys} is to use an RDMA-based meta server to store 
DCT meta. 
Figure~\ref{fig:dc-meta} (a) illustrates the benefit of this design
using the single-connection establishment workload of Figure~\ref{fig:connect-perf} (a).
The baseline (RPC) uses a kernel-space FaSST~\cite{fasst} RPC for the querying.
FaSST is the state-of-the-art RDMA-based RPC that builds on RDMA's unreliable datagram.
It also has no control plane overhead in the kernel because UD is connectionless.
To save CPU resources, 
we only deploy one kernel thread to handle the queries. 
We can see that a meta server design 
achieves an 11.8X better throughput 
and up to 13X query latency compared with RPC. 
The RPC design is bottlenecked by the server CPU for handling DCT queries,
while the RDMA-based meta server bypasses the CPU with one-sided RDMA.

\subsection{Data path performance}
\label{sec:eval-data}

\begin{figure}[!t]
    \begin{minipage}{.48\linewidth}    
    \hspace*{-2mm}
    \includegraphics[scale=1]{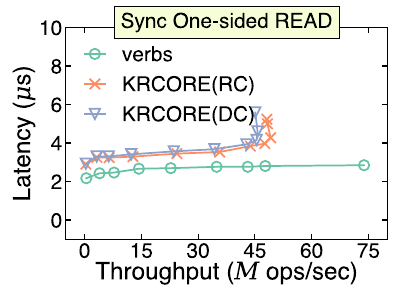} 
    \end{minipage}
    \begin{minipage}{.48\linewidth}    
    \hspace*{-0.5mm}
    \includegraphics[scale=1]{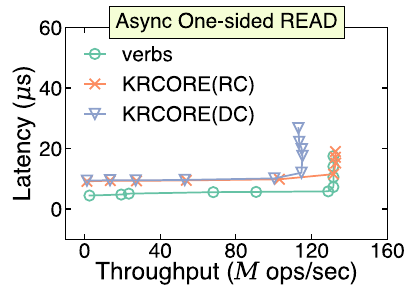} 
    \end{minipage} 

    \begin{minipage}{.48\linewidth}    
        \hspace*{-2mm}
        \includegraphics[scale=1]{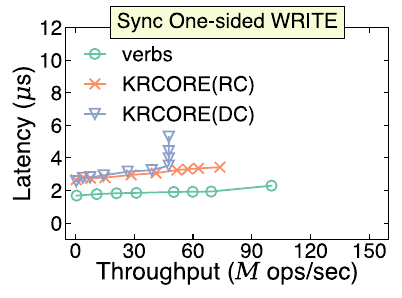} 
    \end{minipage}    
    \begin{minipage}{.48\linewidth}    
        \hspace*{-0.5mm}
        \includegraphics[scale=1]{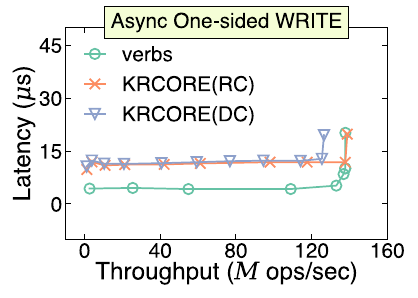} 
    \end{minipage} \\[5pt]
    \begin{minipage}{1\linewidth}        
    \caption{\small{
        The one-sided RDMA performance.
    }}
    \label{fig:eval-one}
    \end{minipage} \\[-10pt]
    \end{figure}    

{\sys} trades data path performance for a faster control plane.
We first use a set of microbenchmarks to evaluate these overheads
using two communication patterns: \textbf{sync} and \textbf{async}. 
In the sync mode, each client issues RDMA requests to one server in a run-to-completion way,
aiming to achieve low latency~\cite{DBLP:conf/nsdi/GaoLTXZPLWLYFZL21,wukong}.
For async, 
each client posts requests in batches to achieve the peak throughput~\cite{drtmh,herd,DBLP:conf/usenix/KaliaKA16}. 
Without explicit mention, the workloads are inbound, 
\ie{multiple clients sending RDMA requests to one server}. 
We reported the aggregated throughput of clients and their average latency.  

\paragraph{One-sided operations. }
Figure~\ref{fig:eval-one} presents the one-sided data path performance of {\sys}
when it virtualizes from DCQP ({\sys}(DC)) and RCQP ({\sys}(RC)), 
and compare them to verbs\footnote{\footnotesize{{\lite}'s 
data path API is different so we compare to it separately. }}.
During the experiment, each client issued 8B random requests to the server, 
and we varied the number of clients from 1 to 240. 

\vspace{0.5mm}
\noindent
(1) Sync. For one-sided RDMA READ in Figure~\ref{fig:eval-one} (a), 
the latency of {\sys} (DC) and (RC) is 27\%--46\% and is 25\%--41\% higher than verbs. 
The additional latency of {\sys} under sync mode is dominated by the system call cost. 
On our hardware, we measure a $\sim$1$\mu$s overhead communicating with the kernel. 
For reference, when using one client, 
the latency of {\sys} (RC) is 3.15$\mu$s, and the verbs is 2.15$\mu$s. 
Another observation is that adopting DCQP has little latency overhead in the sync mode 
as DC reconnection is extremely fast. 
For example, the latency of {\sys} (DC) under one client is 3.24$\mu$s.
The results of one-sided RDMA WRITE in Figure~\ref{fig:eval-one} (c)
are similar to the READ.

\vspace{0.5mm}
\noindent
(2) Async.  For one-sided RDMA READ in Figure~\ref{fig:eval-one} (b), 
{\sys} (RC) can achieve a similar peak throughput as verbs (138M reqs/sec) 
when using 240 clients. 
With the same configuration, {\sys} (DC) is 14\% slower (118 M reqs/sec). 
{\sys} (RC) and verbs are both bottlenecked by the server RNIC, 
while {\sys} (DC) is slower due to extra DCT processing 
at the RNIC. 
For one-sided RDMA WRITE in Figure~\ref{fig:eval-one} (d), 
the results are similar: 
{\sys} (RC) and verbs achieve a peak throughput of 145M reqs/sec 
while {\sys} (DC) is 8.9\% lower (132M reqs/sec). 

\begin{figure}[!t]
    \begin{minipage}{.48\linewidth}    
    \hspace*{-2mm}
    \includegraphics[scale=1]{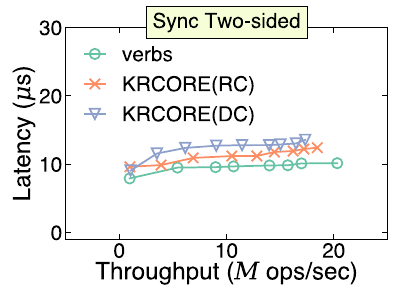} 
    \end{minipage}
    \begin{minipage}{.48\linewidth}    
    \hspace*{-0.5mm}
    \includegraphics[scale=1]{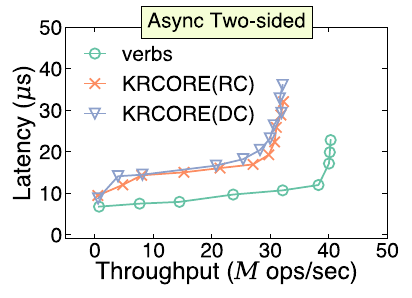} 
    \end{minipage} \\[5pt]
    \begin{minipage}{1\linewidth}
    \caption{\small{
        The two-sided RDMA performance of {\sys}. 
    }}
    \label{fig:eval-two}
    \end{minipage}  \\[-5pt]
    \end{figure}     

\paragraph{Two-sided operations. } 
Figure~\ref{fig:eval-two} presents the two-sided throughput and latency 
of {\sys} w.r.t. to the number of clients (1 to 240).
Each client sends an 8B request to the server in an echo fashion: 
after receiving a request, 
the server will send the request back,
and the client will issue another request after getting the acknowledgment.
The server utilizes all cores (24 threads) to handle these requests. 

\vspace{0.5mm} 
\noindent
(1) Sync. In this mode, the performance comparisons are similar to one-sided RDMA: 
compared with verbs, 
{\sys} (RC) and (DC) have 4--21\% and 14--31\% higher latency, respectively. 
The {\sys} overheads added to two-sided RDMA 
are also dominated by the user-kernel interactions. 
For example, at one client, 
one {\sys} (RC) echo takes 9.6$\mu$s while verbs takes 7.9$\mu$s. 
Compared to one-sided RDMA, the absolute latency gap is larger. 
{\sys} two-sided has an additional system call overhead: 
the server needs to enter the kernel to receive a message.

\vspace{0.5mm}
\noindent
(2) Async. Unlike one-sided RDMA, 
{\sys} cannot achieve the same peak inbound throughput (when using 240 clients)
as verbs for two-sided RDMA: 
it is 20\% slower than verbs: 
which can only achieve 33.7M reqs/sec regardless of RC or DC.
In comparison, verbs can achieve 42.3M reqs/sec. 
The extra bottleneck comes from CPU processing costs at the server 
due to user-kernel interactions.
As a result, {\sys} cannot saturate the RNIC's high performance.
This also explains why {\sys} has a similar performance when using RC and DC.

\paragraph{Effects of zero-copy optimization.} 
We next examine the costs of memory copy---that 
{\sys} uses to dispatch messages between virtual QPs---to the two-sided operations.
We further demonstrate how we mitigate it with a zero-copy protocol (\textsection{\ref{sec:two-sided}}). 
Figure~\ref{fig:dc-meta} (b) shows the two-sided echo latency 
when using one client to communicate with the server w.r.t. the payload size. 
We can see that the memory copy cost is negligible for small transfers (<=16KB)
but is significant for large messages. 
Specifically, when transferring > 16KB messages, the latency of {\sys}
is 1.45--3.1X higher than verbs. 
To this end, the zero-copy optimization ({\sys}+opt) 
reduces the overheads to 0.08-0.23X when transferring >= 16KB messages.

\begin{figure}[!t]
    \begin{minipage}{.48\linewidth}    
            \vspace{2pt}
            \hspace*{-1.5mm}
            \includegraphics[scale=1]{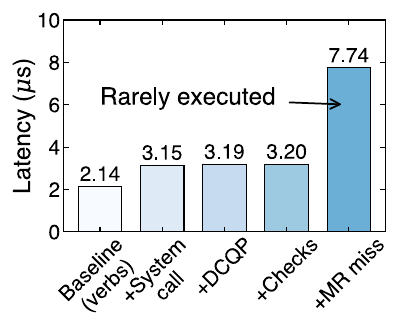} 
            \end{minipage}
            \begin{minipage}{.48\linewidth}    
            \hspace*{-3.5mm}
            \includegraphics[scale=1]{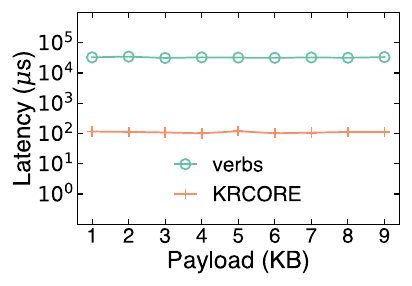} 
            \end{minipage}  
            \begin{minipage}{1\linewidth}
                \caption{\small{ 
                    (a) A factor analysis of the data path cost introduced by {\sys} using one-sided RDMA READ. 
                    (b) The performance of {\sys} in data transfer benchmark of serverless computing. 
        }}                      
    \label{fig:factor}
    \end{minipage} \\[-0pt]
    \end{figure}

\begin{figure}[!t]
    \begin{minipage}{.48\linewidth}    
            \vspace{2pt}
            \hspace*{-3mm}
            \includegraphics[scale=1]{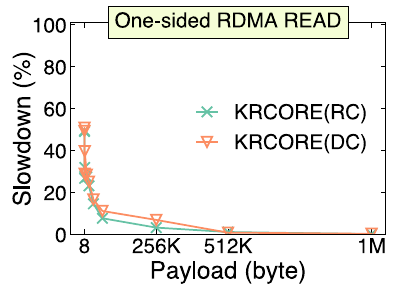} 
            \end{minipage}
            \begin{minipage}{.48\linewidth}    
            \hspace*{-1.5mm}
            \includegraphics[scale=1]{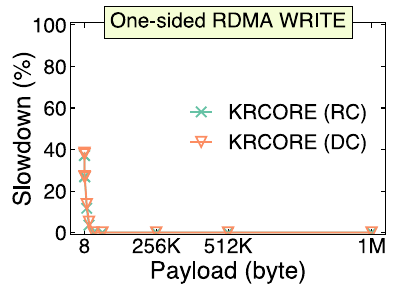} 
            \end{minipage}  \\[2pt]
            \begin{minipage}{1\linewidth}
                \caption{\small{ 
                The slowdown of {\sys} compared to verbs on one-sided RDMA READ (a) and WRITE (b), respectively.
        }}                      
    \label{fig:payload}
    \end{minipage} 
    \end{figure}    

\paragraph{Factor analysis. }
Figure~\ref{fig:factor}(a) conducts a factor analysis 
to show the detailed data path costs of {\sys} in a sync one-sided RDMA READ request. 
The main observations are: 

\vspace{0.5mm}
\noindent
(1) The biggest cost to data path operations is additional RDMA requests to check the MR validity
when the remote MR information is not cached locally (+MR miss, takes 4.5$\mu$s).
Note the checks are rare because {\sys} always caches the checked MR after a miss. 

\vspace{0.5mm}
\noindent 
(2) For normal requests without MR checks, 
system call dominates the overheads (+System call), 
resulting in 1$\mu$s latency increase (3.15$\mu$s vs. 2.14$\mu$s). 
Other costs---including using DCQP (+DCQP) and 
{\sys} check to prevent QP state corruptions (+Checks, see \textsection{\ref{sec:api-data}}) 
are trivial (less than 0.5 $\mu$s). 

\paragraph{Impacts of payload size to one-sided RDMA.}
The overhead of {\sys} becomes smaller for one-sided RDMA with a larger payload,
since transferring data through the network dominates the time. 
Figure~\ref{fig:payload} reports the slowdown compared to verbs on different request payloads.
We measure the latency of sync one-sided RDMA with one client.
For one-sided RDMA READ, the overhead is negligible for larger than 256KB reads (<7\%).
For WRITE, the overhead is negligible for larger than 8KB payloads.

\begin{figure}[!t]
    \begin{minipage}{.48\linewidth}    
            \vspace{2pt}
            \hspace*{-3mm}
            \includegraphics[scale=1]{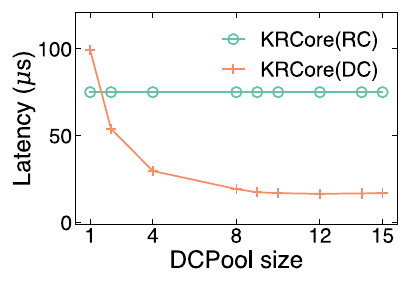} 
            \end{minipage}
            \begin{minipage}{.48\linewidth}    
            \hspace*{-1.5mm}
            \includegraphics[scale=1]{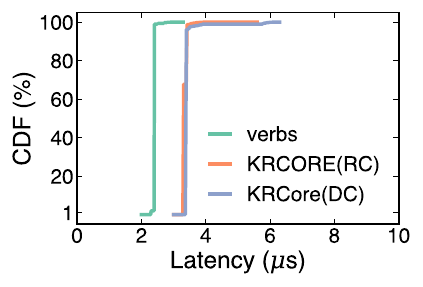}
            \end{minipage}  \\[2pt]
            \begin{minipage}{1\linewidth}
                \caption{\small{ 
                (a) The impacts of DCQP pool size.
                (b) The CDF of latency of sending RDMA requests to different servers.
            }}                      
    \label{fig:dc-pool}
    \end{minipage} \\[-5pt]
    \end{figure}    

\paragraph{Impacts of DCQP pool size. }
A larger DCQP pool is typically better for 
concurrently sending requests to different machines (\textsection{\ref{sec:data-structure}}). 
Figure~\ref{fig:dc-pool} (a) reports the latency when sending 
a batch of 64 one-sided RDMA READs to different targets at one client
with different pool sizes. 
The targets are randomly selected in 10 machines. 
We can see that when the pool only has one DCQP, 
{\sys} (DC) has a 1.32X higher latency (99 vs. 75 $\mu$s) than {\sys} (RC),
since requests to the same QP are processed sequentially with reconnections. 
Increasing the pool size can significantly improve the latency.
Interestingly, when the pool size is larger than 2, 
DC outperforms RC by 28--78\%.
RC needs 64 different connections 
to send these requests, 
and it has to do 63 additional polls than DC.

\paragraph{Tail latency. }
Figure~\ref{fig:dc-pool} (b) reports the tail latency
when using 50 clients sending sync one-sided RDMA READ 
to 5 servers.
Under such a fan-out scenario, 
{\sys} (DC) has a higher tail latency than the others due to 
extra round-trips caused by DC reconnections.
The 99.9\% latency of verbs, {\sys} (RC) and {\sys} (DC)
are 2.8$\mu$s, 3.8$\mu$s and 6$\mu$s, respectively.

\begin{figure}[!t]
            \begin{minipage}{.48\linewidth}    
                    \hspace*{-3mm}
                    \includegraphics[scale=1]{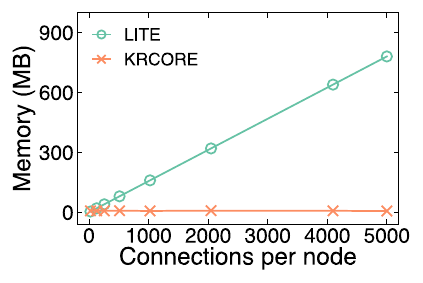} 
                    \end{minipage}
                    \begin{minipage}{.48\linewidth}    
                    \hspace*{0.5mm}
                    \includegraphics[scale=1]{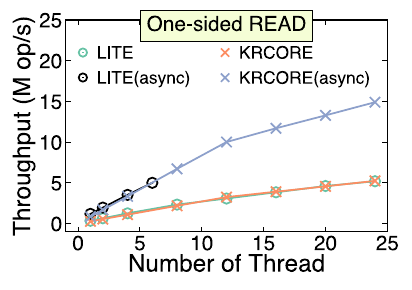} 
                    \end{minipage}  \\[3pt]
                    \begin{minipage}{1\linewidth}
                        \caption{\small{  
                            (a) A comparison of memory usage on connections:
                            {\sys} caches all DCT metadata, while {\lite} caches all RCQPs.
                            (b) A comparison of data path performance when {\sys} uses DCQP. 
                            }}                      
            \label{fig:lite-compare}
            \end{minipage} \\[5pt]
            \end{figure}

\paragraph{Comparison to {\lite}.} 
Finally, 
we show that {\sys} can achieve a similar (or better) data path performance than {\lite} 
with smaller memory usage. 

\vspace{0.8mm}
\noindent 
(1) Memory. 
Figure~\ref{fig:lite-compare} (a) shows the memory used for caching RDMA connections.
In general, {\sys} consumes orders of magnitude smaller memory when 
supporting the same number of connections.
For example,
to maintain 5,000 connections, {\lite} consumes 780MB of memory,
even without counting the memory of message queues (1.5GB if countered). 
In comparison, {\sys} only consumes 6.3MB of memory
because it just maintains a (small) constant number of DCQP (48), 
and each DCT metadata only consumes 12B. 

\vspace{0.8mm}
\noindent 
(2) Performance. 
Figure~\ref{fig:lite-compare} (b) further compares the throughput when 
issuing 64B random one-sided RDMA READ from one node to others. 
We configure both systems to deploy a pool of 32 connections, 
preventing {\lite} from encountering RCQP scalability issues~\cite{fasst}. 
{\sys} uses DCQP for its connections.
For sync, we can see that 
{\sys} is up to 20\% slower than {\lite} due to performance issues of DCQP. 
On the other hand, 
{\sys} achieves a 3X higher peak async throughput (15.6M/sec vs. 5.2M/sec) 
in the async mode. 
{\lite} has a limited peak performance 
because it fails to run with more than 6 threads.
{\lite} doesn't prevent QP queue overflows (see issue \#3 in \textsection{\ref{sec:cost-kernel}}),
so it will trigger QP errors for more than 6 threads.
{\sys} handles overflows well (\textsection{\ref{sec:api-data}}) and can thus,
scale to more threads.

\subsection{Application performance}
\label{sec:eval-app}

\subsubsection{Scaling RACE Hashing}
\label{sec:eval-race}

\paragraph{Overview and setup. }
RACE hashing~\cite{racehashing} is a production RDMA-enabled disaggregated key-value store. 
We chose it as our case study because it requires elastically---a 
demand not commonly found in existing RDMA-based key-value stores.
At a high level, 
RACE separates the storage nodes and computing nodes by RDMA, 
where 
the computing nodes execute key-value store requests
 by issuing one-sided RDMA requests to the storage nodes. 
RACE further allocates computing nodes on-demand 
to cope with various workloads in a resource-efficient way,
where the newly started nodes need dynamically establish RDMA connections to memory nodes. 
To improve performance, 
it embraces a set of low-level RDMA-aware optimizations---\eg{doorbell
batching~\cite{DBLP:conf/usenix/KaliaKA16}} that are tailed to 
RDMA's low-level verbs interface. 

Since RACE is not open-sourced, 
we implement a simplified version atop of verbs, {\lite} and {\sys}, respectively. 
We have calibrated that the performance is close to their reported ones.
For example, 
RACE reports a peak 24M req/sec \texttt{Get} throughput on ConnectX-5 under YCSB-C~\cite{ycsb}.
Our (verbs) version can achieve 27M req/sec with more machines (8 vs. 5) 
on a similar RNIC (ConnectX-4). 

\begin{figure}[!t]
            \centering
            \includegraphics[scale=1]{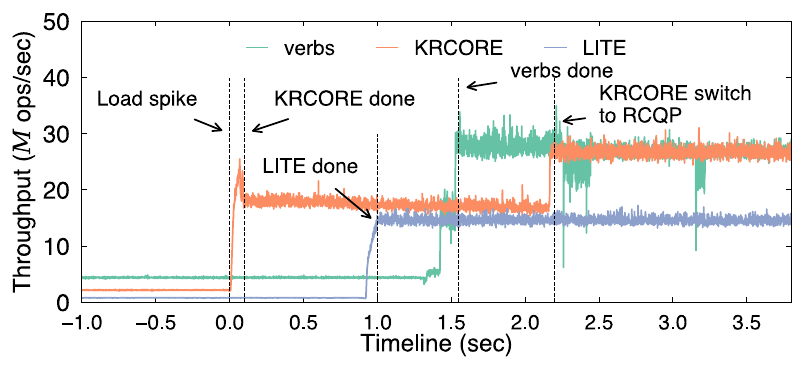}  \\[1pt]
            \begin{minipage}{1\linewidth}
                \caption{\small{  
                Under load spikes, 
                {\sys} can quickly bootstrap computing nodes for RACE Hashing~\cite{racehashing}.
            }}                      
    \label{fig:race-hashing}
    \end{minipage} \\[0pt]
    \end{figure}
	
\paragraph{Performance under load spikes. }
Our evaluating workload contains a load spike
commonly found in real-world applications~\cite{DBLP:conf/sigmetrics/AtikogluXFJP12,DBLP:conf/nsdi/Khandelwal0S16,flashsales}. 
Under spikes, RACE allocates more computing processes to increase performance.
During process startups, {\sys} can reduce its bootstrap time 
thanks to its fast control plane.

Figure~\ref{fig:race-hashing} shows the timelines of RACE 
atop of verbs, LITE and {\sys} under load spikes, respectively. 
The spikes happen at time 0, 
and RACE forks 180 new processors to handle it. 
When using {\sys}, RACE can finish the startup in 244ms, 
83\% and 76\% faster than verbs (1.4 seconds) and {\lite} (1 second), respectively. 
{\sys} is bottlenecked by OS creating worker processors.
On the other hand, {\lite} and verbs are bottlenecked by RDMA's slow control path (\textsection{\ref{sec:bg-issues}}).
A fast boot further reduces the tail latency: 
during time $0$-$3$, {\sys} has a 4.9X lower 99\% latency than verbs.

\paragraph{Benefit of virtualizing a low-level RDMA API. } 
{\sys} virtualizes a low-level RDMA (\eg{\texttt{ibv\_post\_send}}), 
and thus, it can transparently apply existing RDMA-aware optimizations
 (see Issue \#3 in \textsection{\ref{sec:cost-kernel}}).
This leads to better performance of {\sys} on RACE compared to {\lite}:
as shown in Figure~\ref{fig:race-hashing}, 
{\sys} has a 1.73X higher peak throughput (26M reqs/sec vs. 15M reqs/sec) than {\lite} 
after time 3.

\paragraph{Benefit of virtualizing hybrid QPs. }
As shown in Figure~\ref{fig:race-hashing}, 
using RCQP (\eg{after time 3}) brought 1.4X (26M vs. 18M req/sec) throughput improvements to {\sys}, 
achieving a similar performance as verbs (26M reqs/sec). 
This is because RACE issues RDMA requests asynchronously, 
and {\sys}'s RC async peak throughput is similar to verbs (see Figure~\ref{fig:eval-one} (b)).
Further, we can see the overhead of switching from DCQP to RCQP is negligible (at time 2.2).
However,
there is a lag for detecting the switch because 
{\sys} needs time to collect the necessary information to decide which RCQPs to create. 

\subsubsection{Accelerating data transfer in serverless computing}
\label{sec:eval-serverless}

Finally,
we show that {\sys} can improve the communication performance between 
functions in serverless computing.
We use an RDMA-version of data transfer testcase in ServerlessBench~\cite{serverlessbench} (TestCase5), 
a state-of-the-art Serverless benchmark suite.
This testcase measures the data transfer time between two serverless functions.
The experiment runs on Fn~\cite{fn}, a popular open-source serverless platform. 

Figure~\ref{fig:factor} (b) reports the time to pass a message w.r.t. the payload size
when using verbs and {\sys}, respectively.
The receiver function runs in a separate machine using a Docker container after the sender 
finishes execution. 
We use warm start to techniques~\cite{sockatc} to 
reduce the control plane costs of starting containers. 
From the figure we can see
{\sys} reduces the data transfer latency of verbs by 99\% when transferring 1KB to 9KB bytes.
The performance improvements are mainly due to the reduced RDMA control path of {\sys}, 
which we have extensively analyzed in \textsection{\ref{sec:eval-control}}. 

\section{Discussion}
\label{sec:dislim}

\paragraph{Trade-offs of a kernel-space solution. }
{\sys} chooses kernel-space RDMA for a microsecond-scale control plane (5,900X faster than verbs).
Though it retains most benefits of RDMA (\eg{zero-copy}), 
we sacrifice kernel-bypassing benefit and thus,
result in a slower data path (up to 75\% slowdown).
We argue that such cost is acceptable to many elastic applications. 
First, the application usually issues a few networking requests. 
For example, the functions in ServerlessBench~\cite{serverlessbench} 
and SeBS~\cite{DBLP:conf/middleware/CopikKBPH21} only issue one request to read/write remote data on average. 
Second, the control path overhead (ms-scale) 
is commonly orders of magnitude higher than the cumulative data path overhead ($\mu$s-scale), see Figure~\ref{fig:bg-connect-thpt-latency}.
Finally, existing work (\ie{\lite~\cite{tsai2017lite}}) 
also showed that kernel-space RDMA is efficient for many datacenter applications. 

\paragraph{Other RNICs.}
Our analysis focuses on Mellanox ConnectX-4 Infiniband RNIC. 
Nevertheless, we argue the cost is unlikely to reduce 
due to hardware upgrades or different RDMA implementations (\eg{RoCE})
since the cost is dominated by configuring the NIC resources. 
For example, we also evaluate the control path performance on ConnectX-6,
where the user-space driver still takes 17ms for creating and connecting QP, 
similar to the ConnectX-4 we evaluated (15.7ms, see Figure~\ref{fig:bg-connect-thpt-latency}). 

\paragraph{{\sys} in virtualized environments.}
We currently focus on accelerating RDMA control plane with host networking mode. 
Using RDMA in virtual machines or virtualized RDMA network~\cite{freeflow,masq} is also popular in the cloud. 
We believe the principles and methodologies of {\sys} are also applicable in these environments. 
For example, Freeflow~\cite{freeflow} is an RDMA virtualization framework designed for containerized clouds. 
It leverages par-virtualization that intercepts virtualized RDMA requests 
to a software router. 
We can integrate our hybrid connection pool to the router to support a fast control plane atop of it. 
We plan to investigate applying {\sys} in virtualized environments in the future. 

\section{Related Work}
\label{sec:related}

\paragraph{RDMA libraries.}
Many user-space RDMA libraries exist~\cite{DBLP:conf/sigcomm/LiCWBZ19,ucx,rsocket,libvma, DBLP:conf/sosp/ZhangRPONLMLSJP21},
\eg{MPI, UCX~\cite{ucx}, rsocket~\cite{rsocket}}.
They can hardly provide a fast control plane because they all based on verbs.
{\lite}~\cite{tsai2017lite} is the only kernel-space RDMA library and is the closest to our work. 
We have extensively analyzed the issues when deploying {\lite} in elastic computing (\textsection{\ref{sec:cost-kernel}}) 
and how {\sys} addresses them (\textsection{\ref{sec:overview}---\textsection{\ref{sec:design}}}). 

\paragraph{DCT-aware and hybrid-transport systems.} 
Several works used DCT to improve the performance and scalability of RDMA-enabled systems~\cite{DBLP:conf/supercomputer/SubramoniHVCP14, DBLP:conf/saso/ParkY18}. 
Subramoni et al.~\cite{DBLP:conf/supercomputer/SubramoniHVCP14} 
showed that DCT could provide comparable performance to RC while reducing memory consumption for MPI applications. 
Meanwhile, several works leveraged a hybrid-transport design to overcome the shortcoming of a single transport~\cite{DBLP:conf/ipps/KoopJP08, DBLP:conf/ccgrid/JoseSKWWNP12}. 
For instance, Jose et al.~\cite{DBLP:conf/ccgrid/JoseSKWWNP12} utilized UD to reduce the memory consumption of RC in Memcached. 

\paragraph{RDMA-enabled applications.}
{\sys} continues the line of research on accelerating systems with RDMA, 
from key-value stores~\cite{mitchell2013pilaf,xstore,racehashing,herd,farm,cell}, 
far-memory data structures~\cite{DBLP:conf/osdi/RuanSAB20,DBLP:conf/hotos/AguileraKNS19, DBLP:journals/corr/abs-2103-13351}, 
RPC frameworks~\cite{rfp,fasst,DBLP:conf/eurosys/ChenLS19,DBLP:conf/nsdi/KaliaKA19},
replication systems~\cite{DBLP:conf/osdi/AguileraBGMXZ20, DBLP:conf/hpdc/PokeH15, DBLP:conf/cloud/WangJCYC17,DBLP:conf/sigcomm/KimMBZLPRSSS18}, 
distributed transactions~\cite{DBLP:conf/sosp/WeiSCCC15,farmv2,dragojevic2015nocompromises,DBLP:conf/eurosys/ChenWSCC16,drtmh,ford-fast22,DBLP:conf/nsdi/Wei00WGZ21}, 
graphs~\cite{wukong,DBLP:conf/usenix/XieW0C19, DBLP:journals/tpds/YaoCZC22, DBLP:conf/icpp/GuoCZL19} and 
distributed file systems~\cite{DBLP:journals/tos/ZhuCWLS21,DBLP:conf/usenix/LuSCL17,DBLP:conf/fast/YangIS19}, just to name a few. 
Most of these systems do not target elastic computing, 
but we believe there are opportunities for applying them in such a setting. 
In such scenarios, they can benefit from {\sys}. 

\section{Conclusion}
This paper presents {\sys}, 
a $\mu$s-scale RDMA control plane for RDMA-enabled applications that require elasticity. 
By retrofitting RDMA dynamic connected transport 
with kernel-space QP virtualization, 
we show that it is possible to eliminate most RDMA control path costs 
on commodity RNICs.
Meanwhile, the data path costs introduced by {\sys} are 
acceptable for many elastic applications.
Our experimental results confirm the efficacy of {\sys}.

\section{Acknowledgment}
\label{s:ack}
We sincerely thank
the anonymous shepherd and reviewers from USENIX ATC'2022 for their insightful suggestions.
We also thank Dingji Li for discussing how to apply {\sys} to virtual machines, 
Xiating Xie for improving the figures
and Sijie Shen, Zhiyuan Dong, Rongxin Chen and Yuhan Yang for their valuable feedback. 
This work was supported in part by 
the National Key Research \& Development Program of China (No. 2020YFB2104100),
the National Natural Science Foundation of China (No. 61732010, 61925206)
and Shanghai AI Laboratory. 

\balance

\small{
\bibliographystyle{acm}
\bibliography{krdma}
}

\clearpage

\end{document}